\documentclass{emulateapj}
\usepackage{apjfonts}

\slugcomment{Submitted 2009 August 26; accepted 2009 August 30; in original form 2009 June 22}

\shorttitle{Planet migration in evolving disks}
\shortauthors{Alexander \& Armitage}

\begin{document}

\newcommand{\Msunyr}{M$_{\odot}$yr$^{-1}$}
\newcommand{\Mjup}{M$_{\mathrm {Jup}}$}
\newcommand{\Msun}{M$_{\odot}$}

\title{Giant planet migration, disk evolution, and the origin of transitional disks}

\author{Richard~D.~Alexander}
\affil{Sterrewacht Leiden, Universiteit Leiden, Niels Bohrweg 2, 2300 RA, Leiden, the Netherlands}
\author{Philip~J.~Armitage\altaffilmark{1}}
\affil{JILA, 440 UCB, University of Colorado, Boulder, CO 80309-0440}
\altaffiltext{1}{Department of Astrophysical and Planetary Sciences, University of Colorado, Boulder, CO 80309-0391}
\email{rda@strw.leidenuniv.nl}


\begin{abstract}
We present models of giant planet migration in evolving protoplanetary disks.  Our disks evolve subject to viscous transport of angular momentum and photoevaporation, while planets undergo Type II migration.  We use a Monte Carlo approach, running large numbers of models with a range in initial conditions.  We find that relatively simple models can reproduce both the observed radial distribution of extra-solar giant planets, and the lifetimes and accretion histories of protoplanetary disks.  The use of state-of-the-art photoevaporation models results in a degree of coupling between planet formation and disk clearing, which has not been found previously.  Some accretion across planetary orbits is necessary if planets are to survive at radii $\lesssim 1.5$AU, and if planets of Jupiter mass or greater are to survive in our models they must be able to form at late times, when the disk surface density in the formation region is low.  Our model forms two different types of ``transitional'' disks, embedded planets and clearing disks, which show markedly different properties.  We find that the observable properties of these systems are broadly consistent with current observations, and highlight useful observational diagnostics.  We predict that young transition disks are more likely to contain embedded giant planets, while older transition disks are more likely to be undergoing disk clearing.
\end{abstract}



\section{Introduction}\label{sec:intro}
Understanding how planets form has been an active topic of research for centuries, but interest in this subject has increased dramatically since the discovery of the first extra-solar planets \citep{mq95,mb96}.  Over 350 such planets are now known, with a diverse range of properties \citep[e.g.,][]{udry_ppv}.  It was recognized very quickly that many extra-solar planets orbit very close to their parent stars, and that such planets could not have formed at their current locations.  This in turn sparked renewed interest in the established theory of planet migration \citep{gt80,lp86}, in which planets form far from their parent stars and ``migrate'' to smaller radii through the action of tidal torques.  The migration of low-mass planets remains controversial, but so-called Type II migration, which applies to planets sufficiently massive ($\gtrsim0.5$\Mjup) to open gaps in their parent gas disks, is now relatively well understood \citep[see, e.g., the review by][]{pap_ppv}.

In a similar vein, it was discovered in the 1980s that young stars, such as the T Tauri stars (TTs), possess circumstellar disks \citep[e.g.,][]{sb87}.  These disks are of order a percent of the mass of their central star \citep{beckwith90,aw05}, and we now have a well-established evolutionary picture where disk-bearing, classical T Tauri stars (CTTs) evolve into disk-less, weak-lined T Tauri stars (WTTs) on timescales of a few Myr \citep[e.g.,][]{hcga98,haisch01}.  The dominant processes driving (gas) disk evolution are angular momentum transport, and evaporation due to heating by energetic photons \citep[e.g.,][]{holl_ppiv,dull_ppv,rda08}, and these evolving ``protoplanetary'' disks are the sites of planet formation.

Statistically, it is clear that some fraction of observed protoplanetary disks must contain planets.  Identifying these planet-bearing disks is of considerable interest, since doing so would provide evidence as to where and when planets typically form within disks, and discriminate between different models for giant planet formation. Unfortunately, although we have a reasonable idea of what a disk containing a planet would look like (an ordinary disk at large radii, but depleted of gas and dust interior to the planet's orbit), we cannot say that all the disks observed to show such signatures \citep[the so-called transitional disks;][]{strom89,najita07} contain planets.  A significant fraction of transitional disks may instead represent an intermediate stage of disk evolution prior to final disk clearing \citep[e.g.,][]{cieza08}, a process that occurs in the presence or absence of planets. Here, we seek to construct models that include both disk evolution and planet formation.  By comparing the models against both exoplanet and disk evolution statistics we seek to make maximum use of available observational constraints, and thereby predict the observational appearance of planets within evolving disks.

In this paper we present models of giant planet migration in evolving protoplanetary disks.  We restrict ourselves to considering relatively massive planets, $\ge0.5$\Mjup, primarily because searches for exoplanets are presently only complete to around the Jupiter mass level \citep[e.g.,][]{fv05,udry_ppv}.  Consequently, throughout this paper the term ``planet'' always refers to gas giant planets; we make no attempt to model planets of lower mass.  Our models include viscous transport of angular momentum, photoevaporation, and Type II planet migration.  We adopt a Monte Carlo approach, running large numbers of models with a range of initial conditions in order to follow the time evolution of distributions of disk and planet properties.  In \S\ref{sec:models} we present our numerical model, and compare the results to the observed radial distribution of extra-solar planets, and to a large range of observations of protoplanetary disks.  Our results compare favourably with the observed properties of both planets and disks, and we discuss how the observational data can constrain various properties of the model.  This is the first study to model populations of transitional disks theoretically, and in \S\ref{sec:trans} we discuss the transition disk phenomenon.  Our relatively simple model reproduces the known properties of these objects well.  The model produces transition disks via two different mechanisms (gap-opening by planets and disk clearing), and we discuss the relative efficiency of these processes in the models.  We show that observations of transition disk masses and accretion rates remain the most straightforward means of distinguishing between different types of transition disks, and make predictions for future observations of such objects.


\section{Models}\label{sec:models}
\subsection{Planet migration model}\label{sec:model}
In our model, protoplanetary disks evolve due to viscous transport of angular momentum and photoevaporation by the central star.  Planets migrate due to tidal interaction with the disk (in the Type II migration regime), and the disk is also subject to tidal torques from planets.  The coupled evolution of a protoplanetary disk and a planet is described by the equation \citep[e.g.,][]{lp86}
\begin{equation}\label{eq:1ddiff}
\frac{\partial \Sigma}{\partial t} = \frac{1}{R}\frac{\partial}{\partial R}\left[ 3R^{1/2} \frac{\partial}{\partial R}\left(\nu \Sigma R^{1/2}\right) - \frac{2 \Lambda \Sigma R^{3/2}}{(GM_*)^{1/2}}\right] - \dot{\Sigma}_{\mathrm {w}}(R,t) \, .
\end{equation}
Here $\Sigma(R,t)$ is the disk surface density, $t$ is time, $R$ is cylindrical radius, $\nu$ is the kinematic viscosity, and $M_* = 1$\Msun\ is the stellar mass.  The first term on the right-hand side describes ordinary viscous evolution of the disk \citep{lbp74,pringle81}, and the $\dot{\Sigma}_{\mathrm {w}}(R,t)$ term represents the mass-loss due to photoevaporation.    The second term describes how the disk responds to the planetary torque: here $\Lambda(R,a)$ is the rate of specific angular momentum transfer from the planet to the disk.  Following \citet{trilling98} and \citet{armitage02}, for a planet of mass $M_{\mathrm p} = q M_*$ at radius (semi-major axis) $a$ we adopt the following form for $\Lambda$ 
\begin{equation}
\Lambda(R,a) = \left\{ \begin{array}{ll}
- \frac{q^2 GM_*}{2R} \left(\frac{R}{\Delta_{\mathrm p}}\right)^4 & \textrm{if } \, R < a\\
\frac{q^2 GM_*}{2R} \left(\frac{a}{\Delta_{\mathrm p}}\right)^4 & \textrm{if } \,R > a\\
\end{array}\right.
\end{equation}
where
\begin{equation}
\Delta_{\mathrm p} = \textrm{max}(H,|R-a|)
\end{equation}
and $H$ is the disk scale-height.  This form for $\Lambda$ is the same as that used by \citet{lp86}, but modified to give a symmetric treatment inside and outside the planet's orbit.  This transfer of angular momentum causes the planet to migrate at a rate
\begin{equation}
\frac{da}{dt} = - \left(\frac{a}{GM_*}\right) \left(\frac{4\pi}{M_{\mathrm p}}\right) \int R\Lambda \Sigma dR \, .
\end{equation}
This treatment of planet migration is necessarily idealized, but has previously been shown to give results comparable to more sophisticated numerical models \citep[e.g.,][]{tml96}.

The kinematic viscosity $\nu$ governs the transport of angular momentum in the disk, and we adopt an alpha-disk model
\begin{equation}
\nu(R) = \alpha \Omega H^2 \, ,
\end{equation}
where $\alpha$ is the standard \citet{ss73} viscosity parameter and $\Omega(R) = \sqrt{GM_*/R^3}$ is the orbital frequency.  We adopt a scale-height consistent with a flared disk model \citep[e.g.,][]{kh87}
\begin{equation}
H(R) \propto R^p \, ,
\end{equation}
where the power-law index $p=5/4$.  This choice of $p$ gives a viscosity $\nu \propto R$, consistent with high-resolution observations of disk structure \citep[e.g.,][]{aw07,andrews09}.  As in \citet{aa07}, we normalize this relationship so that the disk aspect ratio $H/R=0.0333$ at $R=1$AU.  This parametrization assumes that stellar irradiation dominates over viscous heating, and is thus strictly valid only for accretion rates less than a few times $10^{-8}$\Msunyr\ \citep[e.g.,][]{dalessio99}.  In practice, this means that our model of angular momentum transport in the disk is only physical at times $t \gtrsim 10^{5}$yr.

The rate of mass-loss due to photoevaporation is
\begin{equation}
\dot{\Sigma}_{\mathrm {w}}(R,t) = \left\{ \begin{array}{ll}
\dot{\Sigma}_{\mathrm {diffuse}}(R) & \textrm{if } \, \Sigma_{\mathrm {inner}} > \Sigma_{\mathrm {crit}}\\
\dot{\Sigma}_{\mathrm {direct}}(R) & \textrm{if } \, \Sigma_{\mathrm {inner}} < \Sigma_{\mathrm {crit}}.\\
\end{array} \right.
\end{equation}
The ``diffuse'' profile applies when the inner disk is optically thick to ionizing photons, and was studied in detail by \citet{holl94} and \citet[][see also \citealt{liffman03}]{font04}.  In this case radiation from the star creates a thin ionized layer on the disk surface, with a sound speed $c_s \simeq 10$km s$^{-1}$, and the diffuse (recombination) radiation field from the disk atmosphere is the dominant source of ionizing photons at $\sim$AU radii.  Outside some critical radius the heated layer is unbound and flows as a wind.  The mass-loss profile is strongly concentrated close to the critical radius
\begin{equation}
R_{\mathrm {crit}} \simeq 0.2R_{\mathrm g} \simeq 1.8 \left(\frac{M_*}{1\mathrm M_{\odot}}\right) \mathrm{AU} \, ,
\end{equation}
where $R_{\mathrm g} = GM_*/c_s^2$ is the ``gravitational radius'' defined by \citet{holl94}\footnote{$R_{\mathrm g}$ is found by equating the Keplerian orbital speed with the sound speed of the ionized gas, $c_{\mathrm s}$.}.  The ``direct'' profile applies when the inner disk has been cleared (by either viscous or tidal torques) and is optically thin to ionizing photons.  In this case stellar irradiation ionizes the inner edge of the disk directly, and the disk is cleared from the inside out \citep{acp06a,acp06b}.  We make use of the numerical parametrizations given in the appendix of \citet{aa07}, and switch between the two profiles when the surface density in the inner disk ($R < R_{\mathrm {crit}}$) falls below some critical value $\Sigma_{\mathrm {c}}$.  We set $\Sigma_{\mathrm {c}} = 10^{-5}$g cm$^{-2}$, but note that the results are not very sensitive to the exact value adopted \citep[see discussion in][]{aa07}.  We assume a stellar ionizing flux of $\Phi = 10^{42}$photons per second \citep[e.g.,][]{acp05,ps09,hg09}, which results in a (diffuse) wind rate of $\dot{M}_{\mathrm w} \simeq 4 \times 10^{-10}$\Msunyr.

The initial surface density profile is taken from the similarity solution of the diffusion equation \citep{lbp74,hcga98}:
\begin{equation}
\Sigma(R)=\frac{M_{\mathrm d}}{2\pi R_{\mathrm s} R}\exp(-R/R_{\mathrm s}) \, ,
\end{equation}
where $M_{\mathrm d}$ is the initial disk mass.  The scaling radius $R_{\mathrm s}$ defines the initial disk size, and also sets the viscous time-scale of the disk.  Following \citet{hcga98} we set $R_{\mathrm s} = 10$AU, which results in a viscous time-scale of $t_{\nu} = R_{\mathrm s}^2/3\nu(R_{\mathrm s}) \simeq 5\times10^4$yr for $\alpha=0.01$.


\subsection{Numerical method}
We solve the diffusion equation for the gas surface density (Equation \ref{eq:1ddiff}) numerically using a standard first-order explicit finite difference scheme, on a grid of points equispaced in $R^{1/2}$ \citep[e.g.,][]{pvw86}.  We choose a grid covering the range $[0.04\mathrm {AU},10000\mathrm{AU}]$, and adopt zero-torque boundary conditions (i.e.,~we set $\Sigma=0$ in the boundary cells).  In the absence of a planet the gas disk can be evolved on a fairly coarse grid, but higher resolution is required to maintain accuracy when calculating the effect of the planetary torque.  In order to make efficient use of computing resources we therefore use two resolutions: a low resolution with cell spacing $\Delta R^{1/2} = 0.2$AU$^{1/2}$ and 1000 grid cells, and a high resolution with with $\Delta R^{1/2} = 0.05$AU$^{1/2}$ and 4000 grid cells.  In the absence of a planet the low resolution is used, but we switch to high resolution when a planet is present.  Switching to high resolution is achieved by simple linear interpolation, which is sufficient to conserve mass and angular momentum at very high accuracy.  

The time-step is typically limited by the radial velocity of the gas very close to the location of the planet, and evolving the system on such a short time-step is unnecessary after the planet has opened a gap in the disk.  Consequently we impose a maximum torque (and therefore a maximum gas velocity in the radial direction) of $|\Lambda| \le 0.1 R H \Omega^2$.  In addition, we make no attempt to model the interaction of the planet with the inner edge of the gas disk (which is truncated by the stellar magnetosphere), and simply remove planets when they migrate to radii $a<0.15$AU.  The cumulative numerical errors in the conservation of mass and angular momentum are typically $<0.1$\% over the lifetime of the model.


\subsection{Planetary accretion}
The description of planetary migration above does not allow any material to flow across the gap in the disk induced by the presence of a planet.  However, numerical simulations of the disk-planet interaction show that tidal streams of gas do flow across the gap, allowing accretion on to both the planet and the inner disk to persist after the gap has been opened \citep[e.g.,][]{al96}.  The rate of accretion on to the planet can be parametrized as the fraction $\epsilon$ of the disk accretion rate that would be measured in a steady disk in the absence of a planet.  Simulations show that the efficiency $\epsilon$ varies strongly with the mass of the planet, and can approach (and even exceed) unity for planets which are marginally able to open a gap in the disk \citep{lubow99,dangelo02}.  

\begin{figure}
\includegraphics[angle=270,width=\hsize]{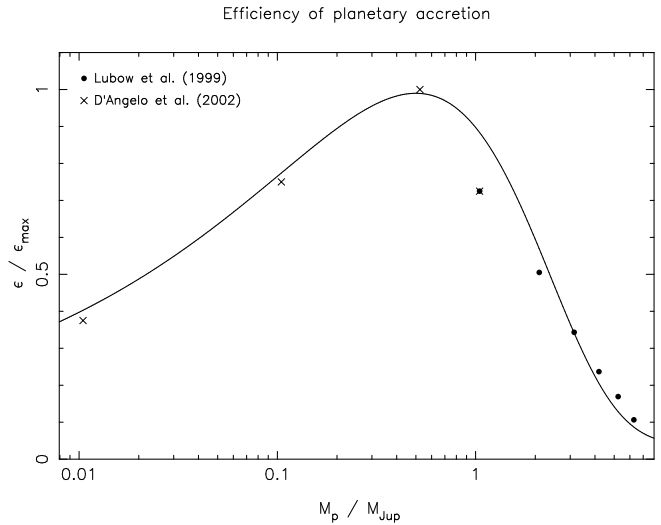}
\caption{Relative efficiency of accretion from the disk on to the planet, plotted as a function of the planet mass.  The crosses show the numerical results from \citet{dangelo02}, and the filled circles the results from \citet{lubow99}.  The solid line shows the fitting formula from \citet{va04} described in Equation \ref{eq:fit}.  The lowest planet mass we consider in our models is 0.5\Mjup.}\label{fig:acc_eff}
\end{figure}

Fig.\ref{fig:acc_eff} shows values of $\epsilon$ obtained from two-dimensional numerical simulations of planets embedded in gaseous disks.  Simulations of low-mass and high-mass planets have been normalized by demanding consistency at $M_{\mathrm p} = 1\mathrm M_{\mathrm {Jup}}$, and further normalized so that the maximum value of the efficiency, $\epsilon_{\mathrm {max}}$, is equal to unity.  In addition, we also show a simple fitting function which provides a good fit to the numerical results \citep{va04}.  The fitting function takes the form
\begin{equation}\label{eq:fit}
\frac{\epsilon(M_{\mathrm p})}{\epsilon_{\mathrm {max}}} =  1.67 \left(\frac{M_{\mathrm p}}{1\mathrm M_{\mathrm {Jup}}}\right)^{1/3} \exp\left(\frac{-M_{\mathrm p}}{1.5\mathrm M_{\mathrm {Jup}}}\right) + 0.04 \, .
\end{equation}
More recent numerical studies suggest that, in addition to allowing accretion on to the planet, tidal streams also permit accretion across the gap from the outer to the inner disk.  Following \citet{lda06}, we define the accretion rate across the gap to be
\begin{equation}
\dot{M}_{\mathrm {inner}} = \frac{1}{1+\epsilon}\dot{M}_{\mathrm p}
\end{equation}
where the accretion rate on to the planet is
\begin{equation}
\dot{M}_{\mathrm p} = \epsilon(M_{\mathrm p}) \dot{M}_{\mathrm {disk}} \, .
\end{equation}
Operationally, we compute the disk accretion rate expected in the absence of a planet, $\dot{M}_{\mathrm {disk}}$ as the accretion rate at three times the radius of the planet:
\begin{equation}
\dot{M}_{\mathrm {disk}} = 3 \pi \nu(3a) \Sigma(3a)
\end{equation}
At each timestep we subtract a mass $dM = dt(\dot{M}_{\mathrm p} + \dot{M}_{\mathrm {inner}})$ from the cell(s) immediately outside the gap, and add the appropriate fractions of this mass to the planet and inner gap edge.  For consistency with the results of \citet{lda06} we adopt $\epsilon_{\mathrm{ max}}=0.5$, but note that our results are not very sensitive to the exact value of $\epsilon_{\mathrm {max}}$.  This procedure does not explicitly conserve angular momentum, as mass is moved from larger to smaller radii without regard for the different angular momenta of these orbits.  What should happen to this ``excess'' angular momentum is not clear \citep{lda06}, but tests show that if all of the excess angular momentum goes to the planet (the extreme case), the migration rate is slowed only at the 1--2\% level.  This is smaller than many of the other uncertainties in the migration model, so we are satisfied that our planetary accretion procedure is robust.


\subsection{Code Tests}\label{sec:tests}
\begin{figure}
\includegraphics[angle=270,width=\hsize]{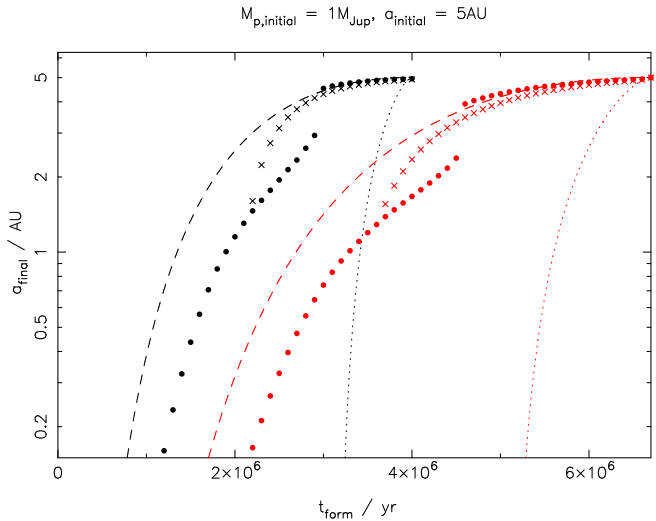}
\caption{Final planet radii as a function of formation time.  The points show our numerical results: black and red represent models with $\alpha=0.01$ and $\alpha=0.003$ respectively.  Circles denote the model where accretion across the planet-induced gap is permitted ($\epsilon_{\mathrm {max}}=0.5$); crosses the model where no gas flows across the gap.  The lines show the corresponding curves from \citet{armitage07}, for both partially (dotted) and fully suppressed (dashed) migration models.  The small discontinuities in the numerical data for the ``flow'' models at final radii $\sim3$AU occur when the planet triggers disk clearing, as discussed in the text. When gas does not flow across the gap this triggering is much more dramatic, and no planets survive at radii $\lesssim 1.5$AU.}\label{fig:a07_comp}
\end{figure}
In order to test the accuracy of our numerical method, we performed a comparison with the semi-analytic results of \citet{armitage07}.  For this test we set the initial disk mass to be $M_d = 10^{-1.5}$M$_{\odot} \simeq 0.0316$\Msun, and ``form'' planets of initial mass $1$\Mjup\ at $a=5$AU.  Figure \ref{fig:a07_comp} shows the final planet radii as a function of formation time, for disks with $\alpha=0.01$ and $\alpha=0.003$ and models with $\epsilon_{\mathrm {max}} = 0.5$ and $\epsilon_{\mathrm {max}} = 0$ (i.e., with and without accretion flow across the planet's orbit).  Also shown are the semi-analytic predictions of \citet{armitage07}, with parameters adjusted to match the disk model used here, for both ``partially suppressed'' and ``fully suppressed'' models \citep[see also][]{sc95,ivanov99}\footnote{Suppression of Type II migration occurs when the planet mass $M_{\mathrm p}$ exceeds the local disk mass $4\pi\Sigma a^2$, as the planet's inertia then causes it to migrate on a time-scale longer than the local viscous time-scale.  The degree of suppression is characterised by the parameter $B = 4\pi\Sigma a^2/M_{\mathrm p}$: for partial suppression the migration rate is reduced by a factor $B^{1/2}$, while in the fully suppressed case it is reduced by a factor $B$.}.  In both cases the numerical results agree very well with the predictions of the fully suppressed migration model, so we are satisfied that our numerical procedure is accurate.  The fact that our models appear consistent with the fully suppressed migration model is not altogether surprising, because by allowing gas to flow across the gap we prevent the pile-up of material near the outer gap edge that drives systems into the partially suppressed regime.

We also see from Fig.\ref{fig:a07_comp} that the inclusion of the direct photoevaporative wind results in a degree of coupling between the formation of planets and the onset of disk clearing.  If a planet suppresses accretion in the inner disk sufficiently, the resulting gap in the disk can become optically thin to ionizing photons.  As a result the wind switches to the direct regime and clears the disk, preventing further planet migration; effectively, the planet triggers disk clearing.  This behaviour is distinct from that considered in previous models \citep[e.g.,][]{armitage02,armitage07}, which assumed that disk clearing was independent of planet formation and migration.  When accretion across the gap is significant the consequences of this are not dramatic (as seen in Fig.\ref{fig:a07_comp}): the location of the small discontinuity in allowed final planet radii differs for different disk models, and when we assume a spread in disk initial conditions the discontinuity is not reflected in the resulting distribution of planet radii.  However, when accretion across the gap is suppressed (i.e., for very low values of $\epsilon$) we see strong coupling between planet formation and disk clearing, with significant implications for the final distribution of planet semi-major axes.  In this case planets can only survive if the outer edge of the gap they induce is outside the critical radius where photoevaporation opens a gap in the disk (approximately 2AU); at smaller radii, the planet is always swept on to the star when the inner disk is cleared.  Consequently, planets can only survive at radii $\lesssim 1.5$AU if there is an accretion flow across the gap.  Such planets are commonly observed, so we conclude that some accretion flow across planetary orbits must occur in real systems.


\subsection{Reference model}\label{sec:ref_model}
In order to study the effects of migration and accretion on the disk and planet properties, we run sets of models in which we randomly sample various model parameters.  We first describe our reference model set, and then a number of further model sets in which we explore the effects of varying different parameters in the model.

In the reference model set the parameters of the disk model are fixed to the values described above, but we allow for a spread in initial disk masses.  We use a log-normal distribution of initial disk masses \citep[see, e.g.,][]{kk09}, with a mean of $\log_{10}(\langle M_{\mathrm d}\rangle / \mathrm M_{\odot}) = -1.5$ and a 3-$\sigma$ spread of 0.5dex.  In the absence of a planet the mean disk model has an initial accretion rate of $\simeq 4 \times 10^{-7}$\Msunyr, starts to undergo photoevaporative clearing after a ``lifetime'' of $\simeq4.2$Myr, and is completely cleared after $\simeq 4.7$Myr.

Not all stars are observed to possess giant planets at AU radii, so we form planets only in a subset of our models: the probability of a giant planet forming in each individual model is 10\% (this value chosen arbitrarily in order to reproduce the observed frequency of giant planets).  In each model which forms a planet, we then allow a single planet of mass $M_{\mathrm p}$ to ``form'' at time $t_{\mathrm p}$ and radius $a_{\mathrm p}$.  In the reference model set we use a constant planet formation radius of $a_{\mathrm p} = 5$AU.  We assign the initial planet masses by randomly sampling a distribution $p(M_{\mathrm p}) \propto 1/M_{\mathrm p}$ \citep[e.g.,][]{marcy08}, over the range 0.5M$_{\mathrm {Jup}} < M_{\mathrm p} < 5.0$\Mjup.  Here the lower limit of 0.5\Mjup\ is approximately the minimum gap-opening mass for our disk model, and the upper limit is chosen, somewhat arbitrarily, near the limit set by the most massive known extra-solar planets.

The time of formation is assigned randomly in the range to 0.25Myr $< t_{\mathrm p} < t_{\mathrm c}$.  The lower limit represents the earliest time at which our disk model is physical (see \S\ref{sec:model} above), and the upper limit is the time at which the wind begins to clear the gas disk.  Following \citet{cc01} and \citet{ruden04} we define this time as 
\begin{equation}
t_{\mathrm c} = \frac{1}{3}t_{\nu} \left(\frac{3 M_{\mathrm d}}{2 t_{\nu} \dot{M}_{\mathrm w}}\right)^{2/3} \, .
\end{equation}
In addition, we limit the maximum value of $t_{\mathrm p}$ such that the planet mass cannot exceed the instantaneous disk mass at $t_{\mathrm p}$.  In practice, however, the wind begins to clear the gas disk when the disk mass reaches $\sim 5$--10\Mjup, so this constraint only applies to the most massive planets.  We ran $N=1000$ randomly realized models: a total of 93 models formed planets.  54 of these planets survived; the remainder were accreted on to the central star.  


\subsection{Results}
\subsubsection{Planet properties}
\begin{figure}
\includegraphics[angle=270,width=\hsize]{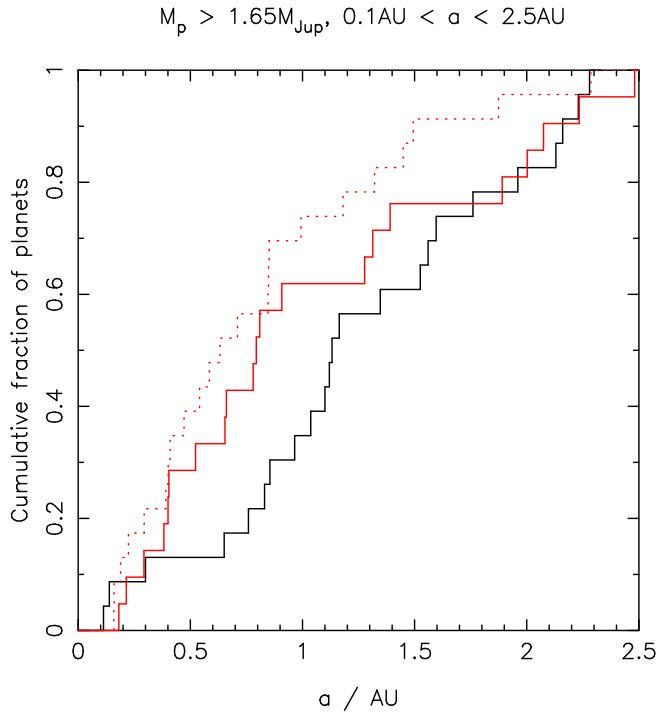}
\caption{Cumulative distributions of planet semi-major axes.  The black line shows the data from \citet{fv05}, cut to be complete in both mass ($M \sin i$) and semi-major axis.  The red line is the corresponding distribution from our reference model.  The dotted red line shows the distribution from a second random realisation of our reference model, and illustrates the typical Poisson errors associated with samples of this size.}\label{fig:ref_distrib}
\end{figure}
A first test of this approach is to reproduce the observed semi-major axis distribution of extra-solar planets.  We follow the approach of \citet{armitage07} in creating a uniformly-selected, complete sample of extra-solar planets, using the data from the Lick radial velocity survey given in \citet{fv05}\footnote{Where possible, we have updated the data from \citet{fv05} using the improved orbital parameters given in \citet{butler06}.}.  These data are complete for Doppler velocities $K > 30$m\ s$^{-1}$ and orbital periods $P < 4$yr.  Consequently, we apply cuts to the sample so that $M \sin i > 1.65$\Mjup\ and $a < 2.5$AU.  Our model makes no attempt to model the survival of ``hot Jupiters'' at small radii, so we also require that $a > 0.1$AU.  This results in a final sample of 23 extra-solar planets (from 850 host stars).  The radial distribution of these planets is shown in Fig.\ref{fig:ref_distrib}, along with the corresponding distribution from our reference model (for the 21 surviving planets with final masses $> 1.65$\Mjup\ and semi-major axes $<2.5$AU).  The two distributions are qualitatively similar, and a Kolmogorov-Smirnov (KS) test fails to reject the (null) hypothesis that that the two data sets are drawn from the same underlying distribution.  (The KS probability is 10\%, with a KS statistic of $D=0.35$.)  The apparent lack of planets around 1.5--2AU is a chance fluctuation, and is not statistically significant (see also Fig.\ref{fig:a07_comp} and the associated discussion in \S\ref{sec:tests}).  Also shown in Fig.\ref{fig:ref_distrib} is the distribution resulting from a second randomly realized set of 1000 models: this time 105 planets formed, and 56 survived.  The null hypothesis is again not strongly rejected by a KS test, but the KS probability of 2\% is somewhat lower than before: this illustrates the typical Poisson errors associated with data sets of this size.  A less conservative cut of the data ($K > 20$m\ s$^{-1}$, $P < 5$yr, $M \sin i > 1.2$\Mjup, $0.1 < a < 3$AU) results in slightly improved number statistics (33 exoplanets), but quantitatively similar distributions of planet semi-major axes.

As a further test, we confirm that the frequency of planets in our models is also consistent with the observed exoplanet statistics.  \citet{cumming08} report frequencies of $2.1\pm0.7$\% for planets with $M \sin i > 2$\Mjup\ and 0.1AU\ $ < a < 2$AU; $0.6\pm0.4$\% for $M \sin i > 2$\Mjup\ and 2AU\ $ < a < 3$AU;  $1.3\pm0.5$\% for 1\Mjup\ $< M \sin i < 2$\Mjup\ and 0.1AU\ $< a < 2$AU; and $0.6\pm0.4$\% for 1\Mjup\ $< M \sin i < 2$\Mjup\ and 2AU\ $< a < 3$AU\footnote{The quoted uncertainties here are purely Poisson errors.  In the \citet{cumming08} sample of 475 stars, the (completeness-corrected) numbers of objects in these four independent bins are 10, 3, 6 \& 3 respectively.}.  The corresponding frequencies for our reference model are 1.3\%, 0.1\%, 2.4\% \& 0.6\% (13, 24, 1 \& 6 planets respectively), which show good agreement with the data.  However, it should be noted that both the absolute frequency and the mass function of planets are primarily determined by their input values, and are essentially free parameters in our model.  We have chosen these in a simple manner in order to be consistent with the observed data, but the radial distribution of planets is the more powerful test of migration physics.  In addition, there is a suggestion that our model over-produces planets of 1--2\Mjup\ at radii $<2$AU.  Given the small numbers of planets this discrepancy is not statistically significant, but it highlights the need for better statistics in this field.  Larger statistical samples of exoplanets will dramatically increase our ability to discriminate between models, and have the potential to provide precise constraints on theories of planet migration \citep[see also][]{armitage07}.

We are therefore confident that our reference model correctly reproduces both the frequency and the radial distribution of known extra-solar giant planets, at least at the $\simeq1$-$\sigma$ level.  The fact that simple models can reproduce this distribution has been noted before \citep[e.g.,][]{armitage07}, but previous such studies have used analytic methods, rather than integrating each model explicitly as we do.  Our method instead ensures that the disk properties (surface density, accretion rate, etc.) of every model are known throughout, and this ``brute force'' method allows us to study the effects of migrating planets on the disk population in a manner that is not possible using more typical population synthesis methods \citep{il04a,il04b,mordasini09a,mordasini09b}.


\subsubsection{Disk properties}
\begin{figure}
\includegraphics[angle=270,width=\hsize]{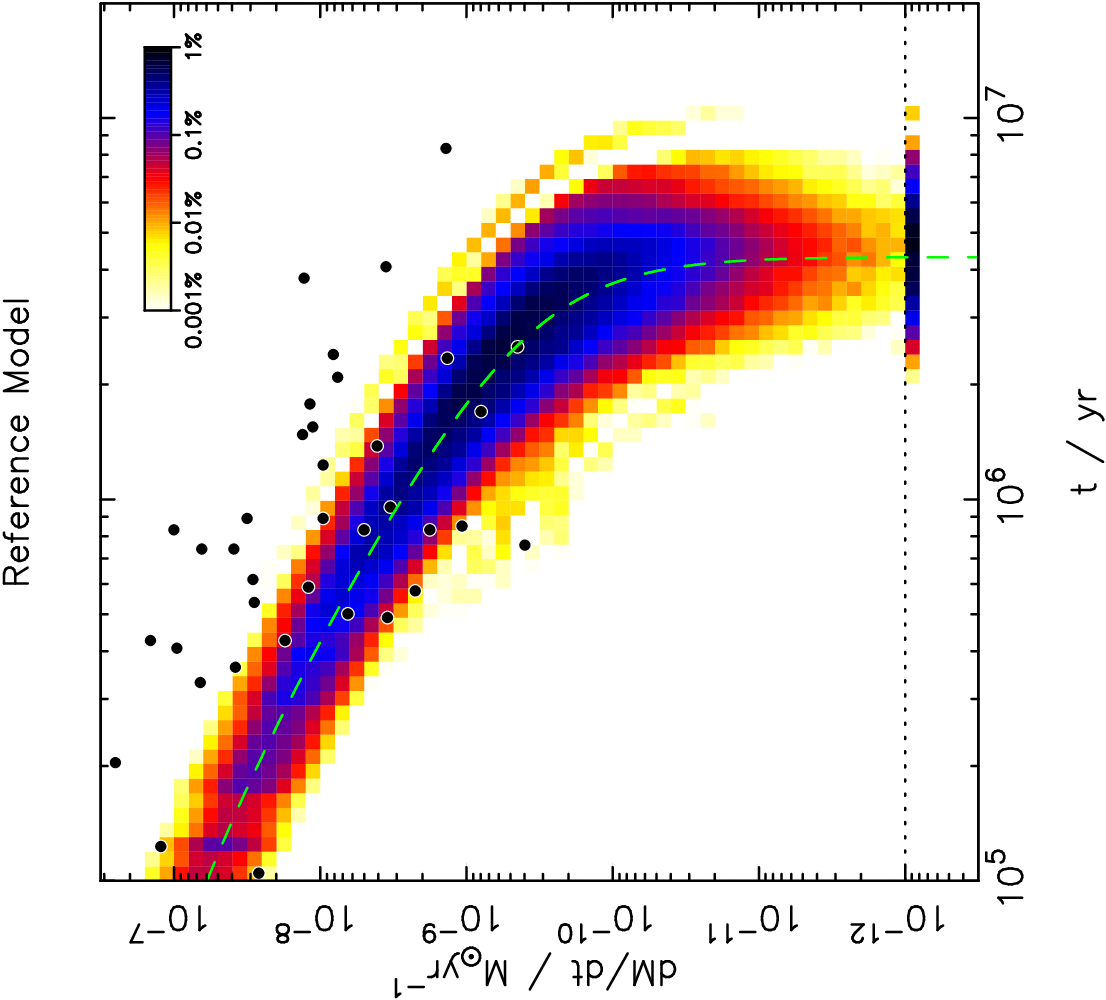}

\vspace*{12pt}

\includegraphics[angle=270,width=\hsize]{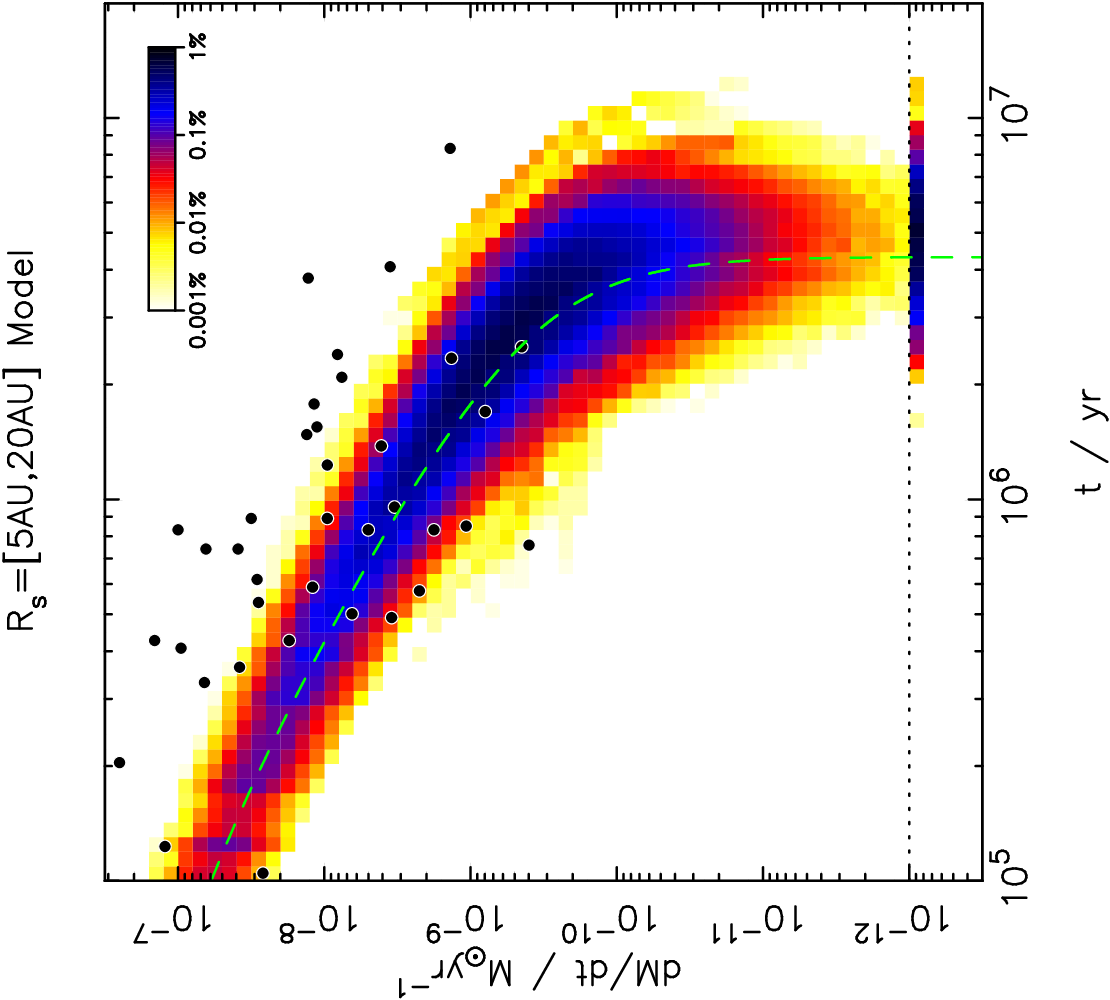}
\caption{Evolution of accretion rate as a function of time in the reference model (upper panel).  The color scale shows the probability of any individual model being found at any given point on the plot.  We regard $10^{-12}$\Msunyr as an arbitrary sensitivity limit, denoted by the dotted line: pixels immediately below this line include all points with accretion rates less than $10^{-12}$\Msunyr, and can be regarded as loci of observational upper limits.  The green dashed line shows the evolution of the median disk model in the absence of a planet.  The data points are taken from \citet{hcga98}.  The lower panel shows the corresponding results for the {\sc scale} model (see \S\ref{sec:params}): the effects of increased dispersion in the disk model are clearly seen.}\label{fig:mdot_time}
\end{figure}
In addition to studying the properties of the migrating planets in our model, we are also able to study the evolution of observable disk properties over the lifetime of the model.  In particular, we are able to follow the accretion rate on to the star (i.e., at the inner boundary) and the disk fraction as functions of time, for our compete set of 1000 disk modes.  The evolution of the accretion rate is shown in Fig.\ref{fig:mdot_time}: the color scale denotes the probability of finding an individual disk at any given position in the $\dot{M}$--$t$ plane\footnote{Note that this probability is computed as a fraction of the total lifetime of our models, and therefore does not account for the duration of the disk-less, WTT phase.}.  The accretion rates decline from a median value of $\simeq 5 \times 10^{-8}$\Msunyr~at $t \simeq 10^5$yr to $\simeq 10^{-10}$\Msunyr~at $t \simeq 4$Myr, at which point the photoevaporative wind becomes dominant and the accretion rate drops precipitously.  There is also significant scatter in the evolution: the shortest disk lifetime is 2.3Myr, while the longest is 10.7Myr.  Much of this scatter is due to the intrinsic dispersion in our disk model, but the effects of migrating planets increase the scatter significantly compared to a disk-only model.  The accretion rates in the model are somewhat lower than in the \citet{hcga98} data, by a factor of $\simeq 3$.  However, this is within the systematic uncertainties associated with both the measured stellar ages and accretion rates, and given the additional uncertainties in our understanding of angular momentum transport in disks this discrepancy is probably not significant.  In addition, the maximum disk masses in the model are $\simeq 0.12$\Msun, with a median value of 0.0045\Msun, in good agreement with the observations of \citet{aw05}.  It has previously been shown that viscous accretion disk models are broadly consistent with the observed evolution of accretion rates in CTTs \citep[e.g.,][]{hcga98,acp03}, and our results re-confirm this conclusion.  

\begin{figure}
\includegraphics[angle=270,width=\hsize]{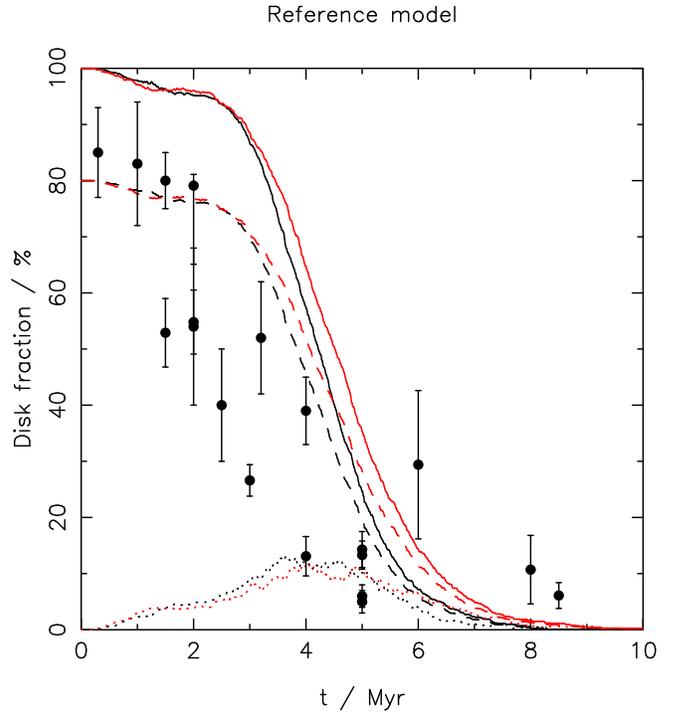}
\caption{Disk fraction as a function of time.  The black lines are from our reference model; the red lines from the {\sc scale} model.  In both cases the solid line is the disk fraction ($N_{\mathrm C}/N_{\mathrm {tot}}$; see \S\ref{sec:trans}), and the dotted line is the fraction of transitional disks ($N_{\mathrm T}/N_{\mathrm {tot}}$; see \S\ref{sec:trans}).  The data points and error bars are the observed disk fractions for a number of nearby star-forming regions, measured primarily by near-infrared excess, compiled by \citet{mamajek09}.  [These data also have significant systematic uncertainties in the derived ages (typically $\pm1$Myr), but for clarity we have omitted these error bars.]  The dashed lines shows the disk fractions multiplied by 0.8, and represent the expected infrared excess fractions once close binaries are taken into account.}\label{fig:disc_frac}
\end{figure}

Fig.\ref{fig:disc_frac} shows the decline in the disk fraction as a function of time.  We define the disk fraction as the fraction of disks of a given age that are ``normal'' viscous disks.  Disks containing migrating planets, and disks which are undergoing photoevaporative clearing, are excluded from this definition and are instead classed as ``transitional'' disks (see \S\ref{sec:trans} below).  This definition should be broadly consistent with previous observational studies of the disk fraction in young star-forming regions \citep[e.g.,][]{haisch01,sicilia06}, which identify disks through near- or mid-infrared excess emission (which is suppressed or absent in both transitional disks and WTTs).  We see that the disk fraction declines from nearly 100\% at $t < 2$Myr to almost zero at $t > 8$Myr, with a median disk lifetime of approximately 4Myr.  This behaviour agrees qualitatively with observational studies \citep[e.g.,][]{haisch01,mamajek09}, but results in a disk lifetimes that are somewhat too long (although, given the large systematic uncertainties in determining the ages of pre-main-sequence stars, it is not clear if this discrepancy is significant).  However, our model set considers only the evolution of disks around single stars, and a significant fraction of young stars are binary or multiple systems.  The evolution of disks in binary systems is complex \citep[e.g.,][]{monin_ppv}, and beyond the scope of this paper, but we note that CTT binaries with small separations ($\lesssim 10$AU) show significantly reduced infrared excesses compared to single CTTs (due to tidal disruption of the inner disk).  Disks in binary systems are interesting in their own right \citep[see, e.g., the discussion in][]{kk09}, but for our purposes these binaries can be regarded as contaminants in photometric studies of disk fractions in young clusters.  Some nearby star-forming regions, such as Taurus-Auriga, have been the subject of detailed multiplicity surveys \citep[e.g.,][]{wg01}, but more distant and more heavily embedded regions have not been similarly studied.  In addition, recent surveys using high-resolution imaging have discovered significant numbers of close binaries even in regions that had previously been well studied \citep{ik08,kraus08}, suggesting that the fraction of stars that exist as binaries with separations of $\sim3$--30AU is not well constrained.  Comparisons to field stars \citep{dm91}, and preliminary results from new observational studies \citep{kraus09}, suggest that the fraction of binaries with such separations is approximately 10--20\%.  We note also that even in the youngest clusters the IR excess fractions rarely exceed 80--85\%, and are inconsistent with 100\% even at ages $\lesssim0.5$Myr \citep[e.g.,][]{haisch01,mamajek09}.  The suggestion that some young stars are born without disks poses serious questions of our understanding of star formation, but a simpler interpretation is that the majority of these ``disk-less'' objects are in fact binary or multiple systems.  If we assume that 20\% of young stars are binaries which do not show strong infrared excesses, then our model agrees very well with the observed data (see Fig.\ref{fig:disc_frac}).  We do not reproduce the long tail of the distribution observed in some older clusters \citep[e.g.,][]{lawson04,sicilia06}, but realistic dispersion in some of the parameters held constant in our models ($\alpha$, $\Phi$, $H/R$), or dispersion in the ages of stars in individual star-forming regions, could easily produce such a tail.  Moreover, some long-lived disks are known to be circumbinary \citep[e.g.,][]{furlan07}, so ``contamination'' by binaries probably plays a role here too.  We are therefore satisfied that our reference model successfully reproduces both the observed distribution of giant extra-solar planets, and the observed properties of disks around young pre-main-sequence stars.


\subsubsection{Effects of model parameters}\label{sec:params}
\begin{table}[!t]
\vspace*{12pt}
\centering
  \begin{tabular}{lccccc}
  \hline\hline
Simulation & $\alpha$ & $\log_{10}(\langle M_{\mathrm d}\rangle / \mathrm M_{\odot})$ & $R_s$/AU & $a_{\mathrm p} /$AU & $M_{\mathrm p}$/\Mjup \\
  \hline
{\sc reference} & 0.01 & $-1.5$ & 10 & 5 & [0.5,5] \\
{\sc radius} & 0.01 & $-1.5$ & 10 & [2,10] & [0.5,5] \\
{\sc radius2} & 0.01 & $-1.5$ & 10 & [5,10] & [0.5,5] \\
{\sc diskmass} & 0.01 & $-1.0$ & 10 & 5 & [0.5,5] \\
{\sc alpha} & 0.003 & $-1.5$ & 10 & 5 & [0.5,5] \\
{\sc scale} & 0.01 & $-1.5$ & [5,20] & 5 & [0.5,5] \\
{\sc fixedmass} & 0.01 & $-1.5$ & 10 & 5 & 0.5 \\
\hline
\end{tabular}
\caption{{\rm List of model sets run.  For each model set 1000 individual models were run, with planets forming in 10\% of the disks.  The values listed for the initial disk mass are the means of the log-normal distributions sampled, as described in the text.  Where the other listed parameters were not fixed, they were randomly drawn from the distributions described in the text [$p(R_s) = \mathrm {constant}$, $p(a_{\mathrm p}) = \mathrm {constant}$, $p(M_{\mathrm p}) \propto 1/M_{\mathrm p}$].}}\label{tab:sims}
\end{table}

In order to study the effect of various parameters on the results of our modeling, we also ran a number of model sets with parameters different from those of the reference model.  In the reference model all planets form at a fixed radius $a_{\mathrm p} = 5$AU.  In reality we expect giant planets to form over a range of radii, so we also considered sets of models where $a_{\mathrm p}$ was assigned randomly in the ranges 2AU$\le a_{\mathrm p}\le10$AU and 5AU$\le a_{\mathrm p}\le 10$AU.  In addition, we consider three other variant model sets: one with $\log_{10}(\langle M_{\mathrm d}\rangle / \mathrm M_{\odot}) = -1.0$; one with $\alpha=0.003$; and one with the scale radius $R_s$ assigned randomly in the range 5AU$\le R_s \le20$AU.  (This last variant has the effect of creating a dispersion in the characteristic viscous scaling time $t_{\nu}$.)  Lastly, we ran a set of models where planets formed with a fixed mass of 0.5\Mjup, rather than the distribution of masses used in the reference model.  The parameters of our model sets are listed in full in Table \ref{tab:sims}. 

The resulting distributions of planet radii unfortunately do not discriminate significantly between the various models.  KS tests fail to reject any of the models which allow for a range of initial planet masses (probabilities in the range 0.5--15\%), and on this basis none of these models are strongly preferred (or disfavoured) with respect to the reference model.  The frequencies of surviving planets vary only weakly between the models, and given that this depends primarily on the input planet formation frequency this also fails to constrain on the model parameters significantly.  The low-viscosity ({\sc alpha}) and high disk mass ({\sc diskmass}) models are somewhat disfavoured due to their long disk lifetimes (mean disk lifetimes of 7.4 \& 9.8Myr respectively), but given the intrinsic uncertainties in our understanding of angular momentum transport in disks this is also not especially significant.  Similarly, the {\sc scale} model, in which the disk scale radius $R_s$ varies, is weakly preferred over the reference model, due to the increased dispersion in the resulting disk lifetimes and accretion rates (see Fig.\ref{fig:mdot_time}).  Allowing for ranges in planet formation radii does not significantly alter the distribution of planets in the ``migration zone'' ($\lesssim 3$AU), but does give rise to significant differences in the distribution of planets at larger radii ($\sim5$--10AU).  When the radial velocity surveys for planets become complete to larger radii (longer orbital periods), they will provide stronger constraints on the radii at which giant planets form \citep[see also][]{armitage07}.

The one model which is strongly rejected is that in which planets form with a fixed mass of 0.5\Mjup\ (model {\sc fixedmass}).  In this model dispersion in planet masses is solely due to differences in the accretion history of planets during the migration phase, and consequently the final planet masses correlate strongly with radius.  Planets which spend a longer time migrating accrete more gas, so the most massive planets are always found at small radii; such a correlation is not observed in exoplanet surveys.  Moreover, planetary accretion is not fast enough to account for the observed range in planet masses: the most massive surviving planets have masses of $\simeq 1.5$\Mjup.  We are therefore able to reject this model at high confidence, and conclude that giant planets must enter the Type II migration regime with a large range of masses.


\subsection{Discussion}
A critical feature of our migration model is the maximum time at which planets are allowed to form, $t_c$.  Planets which form at late times are only able to migrate a limited distance before the disk is cleared (see Fig.\ref{fig:a07_comp}), so the latest time at which planets can form has important implications for the resulting distribution of planets, especially close to the formation radius $a_{\mathrm p}$.  Variations in $t_c$ do not strongly affect the distribution of planets in the reference model at radii $\lesssim 3$AU, but changing $t_c$ by as little as 10\% can result in 3-$\sigma$ changes in the distribution at larger radii.  We therefore attach limited significance to our predicted distributions of planets in the range $\sim$3--10AU.  However, we note that even the formation of the observed population of giant planets (in the migration zone) requires planets to form at relatively late times, $\gtrsim$1--2Myr (see Fig.\ref{fig:a07_comp}).  This result is predicated on the assumption that protoplanetary disks accrete in a manner consistent with our viscous accretion model.  Adopting lower values of $\alpha$ reduces the efficiency of Type II migration somewhat, but one cannot adopt arbitrarily low values and still reproduce the observed accretion rates.  As long as disks are accreting viscously at the observed rates, it is difficult to avoid the conclusion that giant planets must be able to form late.  At these times the disk masses in our models are low, $\lesssim0.01$\Msun, with surface densities at 5--10AU of $\lesssim10$g cm$^{-2}$.  This is at least an order of magnitude smaller than the surface density in the canonical Minimum Mass Solar Nebula \citep{weidenschilling77,hayashi81}, which is treated as a fiducial value in many calculations of planet formation.  Forming planets in disks with such low surface densities may be challenging for modern theories of planet formation \citep[e.g.,][]{johansen07}.

A further limitation of our models is that we consider only the formation of one planet per disk, while in reality many planetary systems contain multiple planets.  In such systems planet-planet interactions can be important, and can modify the extra-solar planet distribution after the dispersal of the gas disk.  Indeed, it seems likely that planet-planet scattering is responsible for the observed distribution of exoplanet eccentricities \citep{jt08,chatterjee08}.  In this scenario many planets in the migration zone undergo additional migration after the gas disk is cleared, and this process causes the innermost planet to migrate by an amount that depends, on average, on the number of giant planets present at the end of the disk lifetime.  If this number is typically small (2--3) scattering will result in a modest re-mapping of our planet distributions to smaller radii, but will not remove the need for the distribution of planets at small semi-major axes to be primarily established through disk migration.  The shape of the distribution would also change if the scattering properties are not scale-free in radius.  Considering these effects would be necessary in precision tests of data against theory, but is not warranted with the limited suitable data samples available at present.


\section{Transitional disks}\label{sec:trans}
Since their discovery by \citet{strom89}, the so-called ``transitional disks'' have been thought to represent a crucial step in the evolution of planet-forming disks around young stars.  These disks are characterized by reduced emission in the infrared but longer wavelength emission consistent with normal CTT disks, and this is generally attributed to some degree of inner disk clearing.  The fraction of disks which appear transitional is small, typically 5--10\% \citep{skrutskie90,kh95,aw05}, which leads to the conclusion that the transitional phase is short-lived \citep{sp95,ww96}.  Unfortunately this also means that the sample of well-studied transitional disks is small, and this small sample shows considerable diversity in disk properties.  A detailed understanding of this important phase of disk evolution has thus so far remained elusive \citep[see the discussion in][]{rda08}.

Theoretically, a number of different physical processes are expected to give rise to disks with ``transitional'' spectral energy distributions: planets \citep{rice03,quillen04}, dust evolution \citep{dd05,krauss07}, disk clearing \citep{acp06b,cmc07} and the presence of companions \citep{jm97,ik08} probably all play a role.  However, recent observational studies, especially those with the {\it Spitzer Space Telescope}, have led to a dramatic increase in the number of known transitional disks, resulting in the first demographic studies of their properties.  \citet{najita07} identified a sample of 12 transitional disks in the Taurus-Auriga cloud, using 5--30$\mu$m data from the {\it Spitzer} spectroscopic survey of \citet{furlan06}.  All of the 8 single stars in their sample are actively accreting, but the accretion rates for the transitional disks were found to be, on average, an order of magnitude lower than for CTTs with similar disk masses.  \citet{najita07} concluded that partial inner disk clearing by embedded planets was the most probable explanation for this result, but noted that other explanations (notably dust settling or growth) could not be ruled out.  By contrast, \citet{cieza08} identified a sample of 26 transitional disks in a number of nearby star-forming clouds, using different selection criteria based on observations across a wider range in wavelengths.  They found that inner disk clearing was associated with significant depletion of the (outer) disk mass, and concluded that disk evolution (presumably due to viscosity and/or photoevaporation) was the most probable explanation.  These contrasting results suggest that selection biases still dominate these relatively small samples, but also suggest that more than one physical mechanism is responsible for the systems that are broadly classed as ``transitional''.  This view is further supported by the recent results of \citet{salyk09}, who used observations of CO emission lines to divide a sample of 14 transition disks into ``cleared'' and ``partially depleted'' inner disks.

A key issue in such studies is how transitional disks are defined.  A broad definition, such as that used by \citet{najita07}, encompasses settled dust disks, disks where significant grain growth has occurred, disks with inner holes, disks with embedded planets, and some binaries.  A more strict definition, such as that proposed by \citet{rda08}, limits the sample to objects with partially or fully cleared inner holes, and thus only selects objects whose gas disks have undergone significant evolution or perturbation.  In our models we can identify, and distinguish, two different types of ``transitional'' disk: disks with holes or gaps due to embedded planets, and disks which are being cleared (through the combined action of viscosity, photoevaporation, and possibly planetary torques).  \citet{rice06} demonstrated that a planet which opens a gap in a disk reduces the dust-to-gas ratio in the inner disk, resulting in a corresponding suppression of infrared emission.  As we only consider planets which are massive enough to open such a gap, all planet-bearing disks in our models are likely to be identified as transitional.  In addition, any disk which has evolved to be optically thin in the inner disk at infrared wavelengths but remains optically thick at larger radii (i.e., any disk undergoing inside-out clearing) is likely to be classed as transitional.  Consequently, in our models we identify disks as transitional if 

\begin{center}
$\Sigma(1\mathrm {AU}) < 10^{-2}$g\ cm$^{-2}$ AND $\Sigma(R_h) > 10^{-4}$g\ cm$^{-2}$

OR

the disk contains a planet.  
\end{center}
Note, however, that disks identified in this manner represent only a subset of the observed samples of transitional disks, which are generally more broadly defined (as discussed above and in \S\ref{sec:discussion}).  

In our models the disks are able to spread to arbitrarily large radii, and many of our disks expand to radii of order 1000AU.  However, such large disks are not commonly observed \citep[e.g.,][]{aw07,andrews09}, and in reality the outer edges of disks are likely to be truncated by a variety of different physical processes (such as photoevaporation by non-ionizing radiation, or tidal interactions with other stars).  This simplification does not have a strong effect on the global evolution of our models, but does cause problems in defining when disks have been cleared.  We define all disks which have inner holes larger than some critical radius $R_h$ to be ``cleared''.  Initially we adopt $R_h = 100$AU, but we also investigate the effect of varying this parameter on our results.  

For clarity we define the following quantities in each set of models, which vary as functions of time:
\begin{itemize}
\item[$N_{\mathrm {tot}}$] The total number of stars (always 1000 in our models).
\item[$N_{\mathrm C}$] The number of stars with normal, non-planet-bearing disks, analogous to the number of CTTs (or Class II sources).
\item[$N_{\mathrm T}$] The number of stars with transitional disks (according to the above criteria).
\item[$N_{\mathrm P}$] The number of stars with planet-bearing disks.  These represent a subset of the more broadly-defined transitional disks.  (Note that stars with planets but no disk are classified as disk-less.)
\item[$N_{\mathrm W}$] The number of stars with disks cleared to beyond $R_h$, analogous to the number of WTTs (or Class III sources).
\end{itemize}
By construction, $N_{\mathrm C}+N_{\mathrm T}+N_{\mathrm W} = N_{\mathrm {tot}}$.  We note in passing that the term transitional is rather misleading in this context, as many of our disks evolve from a ``transitional'', planet-bearing phase back into a normal CTT phase before they are finally cleared.


\subsection{Number statistics}\label{sec:num_stats}
By adopting the definitions above, we are able to identify a subset of transitional disks in our model sets that is broadly consistent with observationally-defined samples of transitional disks.  The fraction of transitional disks ($N_{\mathrm T}/N_{\mathrm {tot}}$) in the reference model varies from 0--13\%, as seen in Fig.\ref{fig:disc_frac}.  Varying the value of $R_h$ has only a small effect on this result: for $R_h=30$AU the peak transition disk fraction is 10\%, while for $R_h = 300$AU it rises to 17\%.  The transition fraction is initially small, rises to a peak at approximately the median disk lifetime, and then declines.  The increase in the transition disk fraction with time can be understood in terms of the increasing migration time-scale (due to increased suppression of migration with declining disk mass), and also the increasing incidence of clearing discs at later times; the decline at late times is imposed by overall decrease number in the number of disk-bearing systems.  However, some aspects of the shape of this curve are artefacts of our model: in particular, the lack of transition disks at early times is in part due to the fact that we do not form any planets at $t < 0.25$Myr.  In addition, we assume zero dispersion in the age of our populations, while real clusters may have significant age spreads.  We therefore urge caution when comparing our results to observations of very young clusters.

\begin{figure}
\includegraphics[angle=270,width=\hsize]{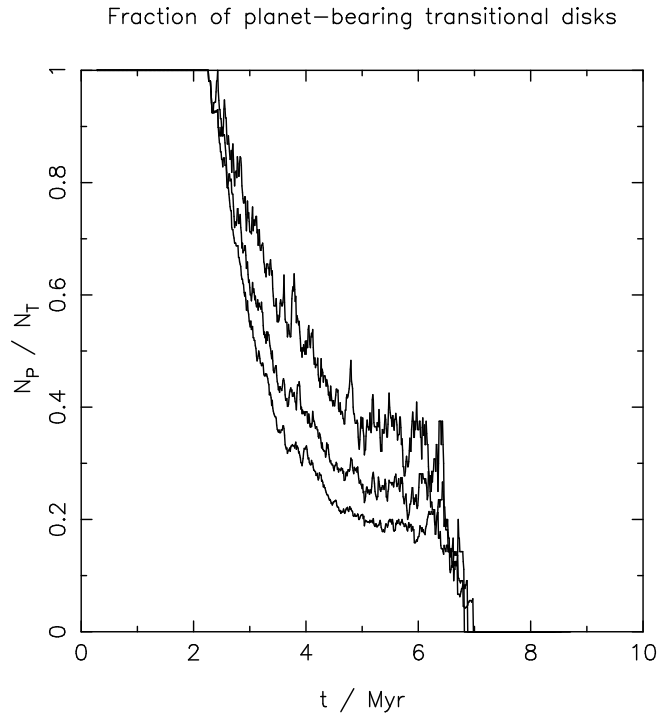}
\caption{The fraction of transitional disks in the reference model set which contain planets ($N_{\mathrm P}/N_{\mathrm T}$), plotted as a function of time.  From top to bottom the three curves are calculated for $R_h=30$, 100 \& 300AU respectively: selecting transitional disks with larger hole sizes increases the fraction of clearing disks.  At early times all of the transitional disks contain embedded planets.}
\label{fig:planet_fraction}
\end{figure}

Fig.\ref{fig:planet_fraction} shows the fraction of transitional disks which possess planets ($N_{\mathrm P}/N_{\mathrm T}$), as a function of time, for three different values of $R_h$ (30, 100 and 300AU).  Larger values of $R_h$ naturally lower the ratio $N_{\mathrm P}/N_{\mathrm T}$ (because all disks with such large holes must be clearing), but the same general trend is seen in all the models.  At early times, the disk accretion rates are too high for photoevaporation to be important, so essentially all of the transitional disks with ages $\lesssim 2$Myr are planet-bearing.  As the population evolves photoevaporation becomes important for a progressively larger fraction of disks, so the number of clearing disks increases and the ratio $N_{\mathrm P}/N_{\mathrm T}$ declines.  Here the time-scale of $\sim2$Myr is simply the point at which the first of our disks (those with the lowest initial masses) reach the low, $\sim10^{-10}$\Msunyr\ accretion rates where photoevaporation becomes important.  The absolute time-scales in our model are set by the disk viscosity and initial conditions, which are essentially free parameters chosen to match the observed constraints on disk evolution (as discussed in Section \S\ref{sec:ref_model}).  We choose these parameters such that the median disk lifetime is $\simeq$4Myr, and the time at which the first clearing disks appear is determined by the dispersion in disk lifetimes.  In the reference model set this is set by the dispersion in initial disk masses, and the first clearing disks appear after $\sim2$Myr; in the {\sc scale} model set there is an additional dispersion in the viscous scaling time, and the first clearing disks appear slightly earlier ($\sim1.8$Myr).  However, unless the dispersion in disk lifetimes is very large (of order the median lifetime), it is very unlikely that disk clearing will be significant at ages $\lesssim 1$--2Myr\footnote{If, however, disk clearing was to begin at higher accretion rates \citep[as suggested by][]{ercolano09b,owen09}, we may see qualitatively different behavior.}.  These results suggest that younger transitional disks may be the most promising candidates for hosting embedded planets (at least in a statistical sense).  They also suggest that longer wavelength observations, which are sensitive to larger hole sizes, should preferentially detect clearing transitional disks, and that new facilities (such as the {\it Herschel Space Observatory}) may discover large numbers of these objects.

\begin{figure}
\includegraphics[angle=270,width=\hsize]{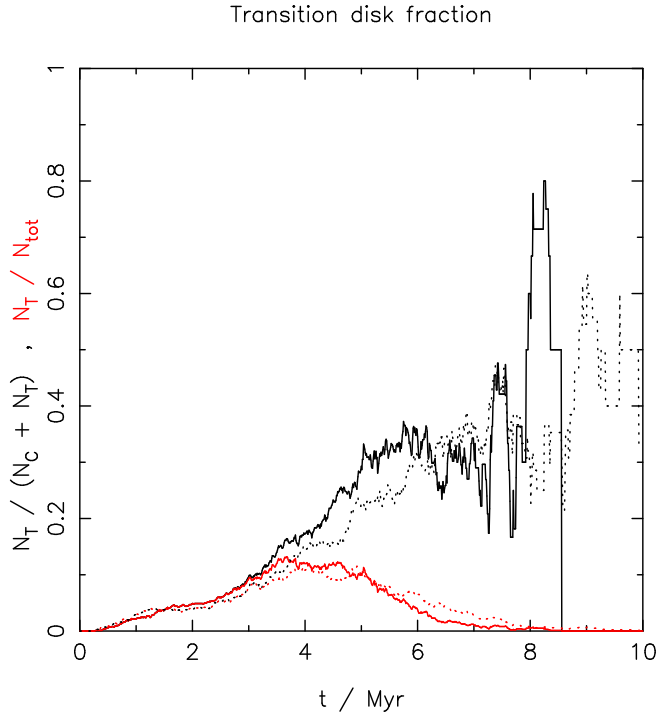}
\caption{The fraction of disks that are transitional [$N_{\mathrm T}/(N_{\mathrm C}+N_{\mathrm T})$; black lines], plotted as a function of time (for $R_h=100$AU).  The solid line shows the results from the {\sc reference} model; the dotted line the {\sc scale} model.  At late times, when few disks remain, this number approaches unity.  Also shown for comparison are the ``true'' transition disk fractions $N_{\mathrm T}/N_{\mathrm {tot}}$, for the {\sc reference} (solid red line) and {\sc scale} (dashed red line) models.}\label{fig:nt_nc}
\end{figure}

Fig.\ref{fig:nt_nc} plots the fraction of disks that are transitional [i.e., $N_{\mathrm T}/(N_{\mathrm C}+N_{\mathrm T})$] as a function of time, for the {\sc reference} and {\sc scale} models.  At early times this ``transition fraction'' is small, but as the total disk fraction declines $N_{\mathrm T}/(N_{\mathrm C}+N_{\mathrm T})$ increases to around 30\%, and approaches unity at very late times (when only a handful of disks remain, most of which are clearing).  In addition, the small denominator causes this ratio to become highly stochastic at last times.  We stress that this behavior is not inconsistent with the rapid disk clearing seen in our models \citep[as claimed by, e.g.,][]{currie09}, but is instead a natural consequence of a rapid transition in a population of stars with a plausible dispersion in disk lifetime.  In addition, we note that our models somewhat under-estimate the fraction of transitional disks, as we do not consider processes, such as dust settling or terrestrial planet formation, that may lead to the appearance of ``homologously depleted'' transitional disks \citep[e.g.,][]{wood02}.  Consequently at late times, when only a few disks remain, many, if not most of them will be transitional.  We therefore urge against using the ratio $N_{\mathrm T}/(N_{\mathrm C}+N_{\mathrm T})$ (or, similarly, $N_{\mathrm T}/N_{\mathrm C}$) as a measure of the transition disk fraction, especially in clusters older than a few Myr (where the disk fraction $N_{\mathrm C}/N_{\mathrm {tot}}$ is small).  Although it is observationally less convenient (as it requires an accurate census of the number of disk-less stars), the ratio $N_{\mathrm T}/N_{\mathrm {tot}}$ is a more robust statistical measure of the duration of the transition disk phase.


\subsection{Transition disk properties}
Recently, \citet{aa07} and \citet{najita07} independently proposed that statistical studies of the masses and accretion rates of transitional disks could provide insight into the physical nature of these systems.  Embedded giant planets are expected to suppress accretion without significantly altering the disk mass, while photoevaporative clearing requires significant evolution of the entire disk before the accretion rate falls to a low enough value for a gap to open.  However, the theoretical arguments used in these papers were highly idealized, and did not consider the time-dependence of the planet-disk interaction.  Our models allow us to study these processes more fully, and make detailed predictions about the properties of (some) observable transitional disks.

\begin{figure}
\includegraphics[angle=270,width=\hsize]{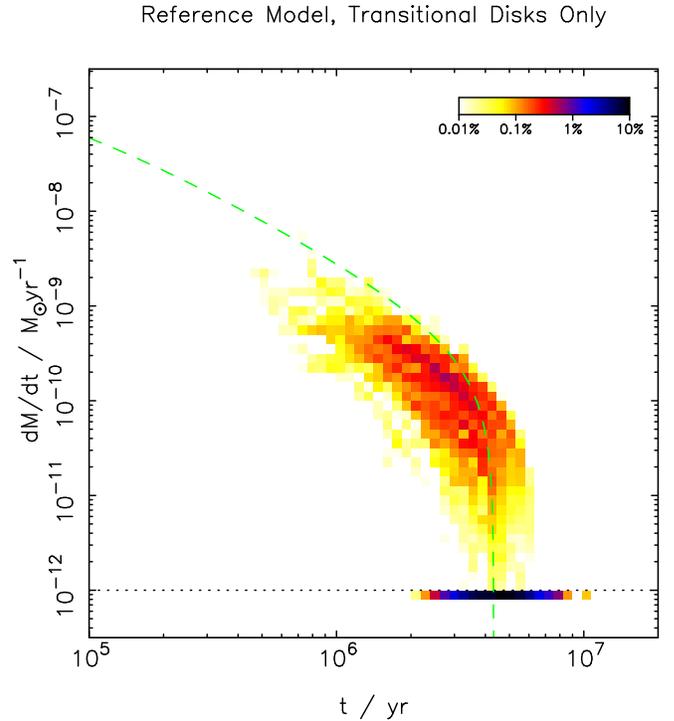}
\caption{As Fig.\ref{fig:mdot_time}, but for transitional disks only.  The green dashed line again shows the evolution of the median disk model.  The accretion rates for accreting transitional disks are suppressed by around an order of magnitude relative to the disk population as a whole.}\label{fig:mdot_t_trans}
\end{figure}

Fig.\ref{fig:mdot_t_trans} shows the evolution of accretion rates in the reference model for disks that are identified as transitional.  At ages $<2$Myr (see discussion in \S\ref{sec:num_stats} above) essentially all of the transitional disks are accreting, but at later times there are two coeval populations of transitional disks: accreting, planet-bearing disks and non-accreting, clearing disks (a few of which also contain planets).  In the subset of accreting transitional disks the accretion rates show considerable scatter, and the median is suppressed by approximately a factor of 10 with respect to the CTT population.  This is consistent with the results of \citet{najita07}, and supports their suggestion that the transitional disks in their sample possess embedded planets.

\begin{figure}
\includegraphics[angle=270,width=\hsize]{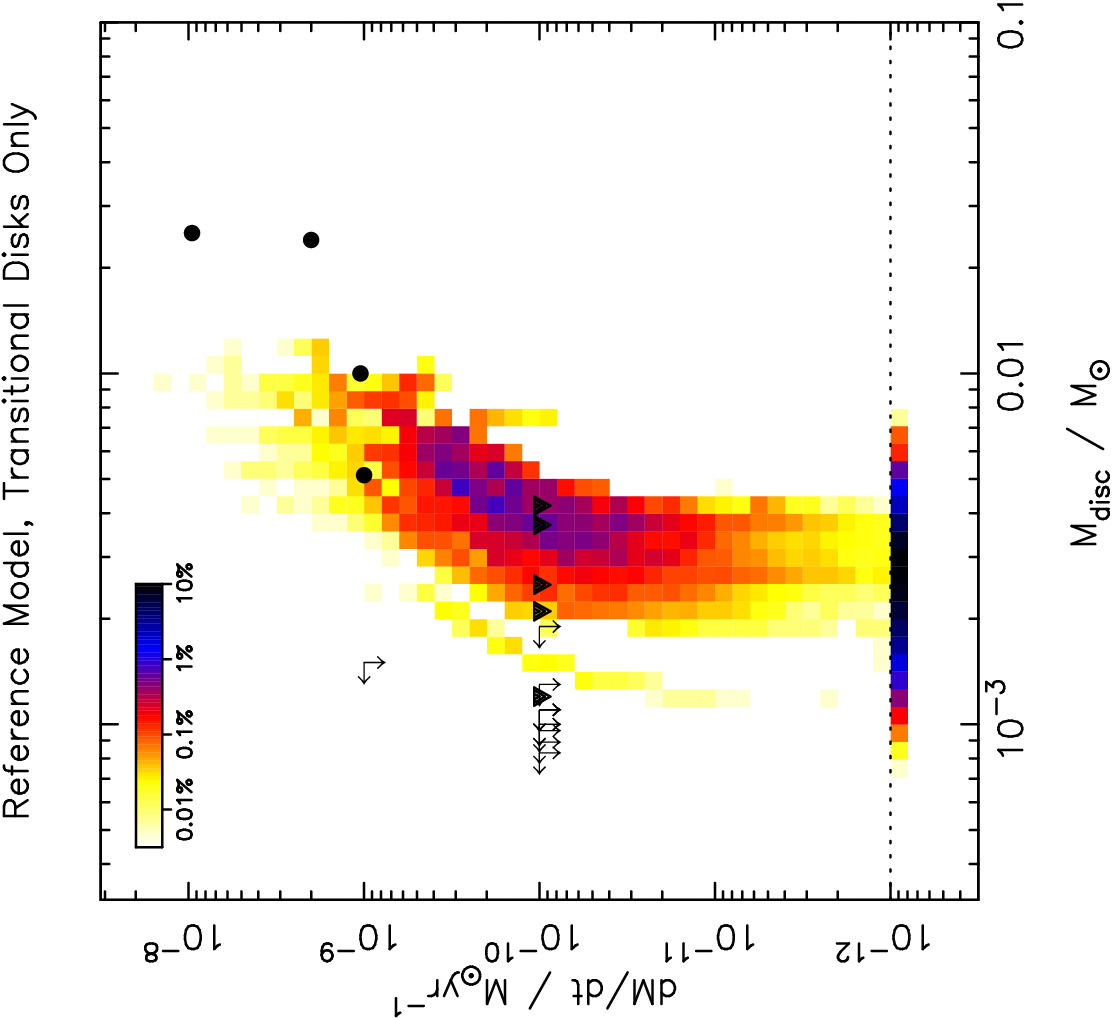}

\vspace*{12pt}

\includegraphics[angle=270,width=\hsize]{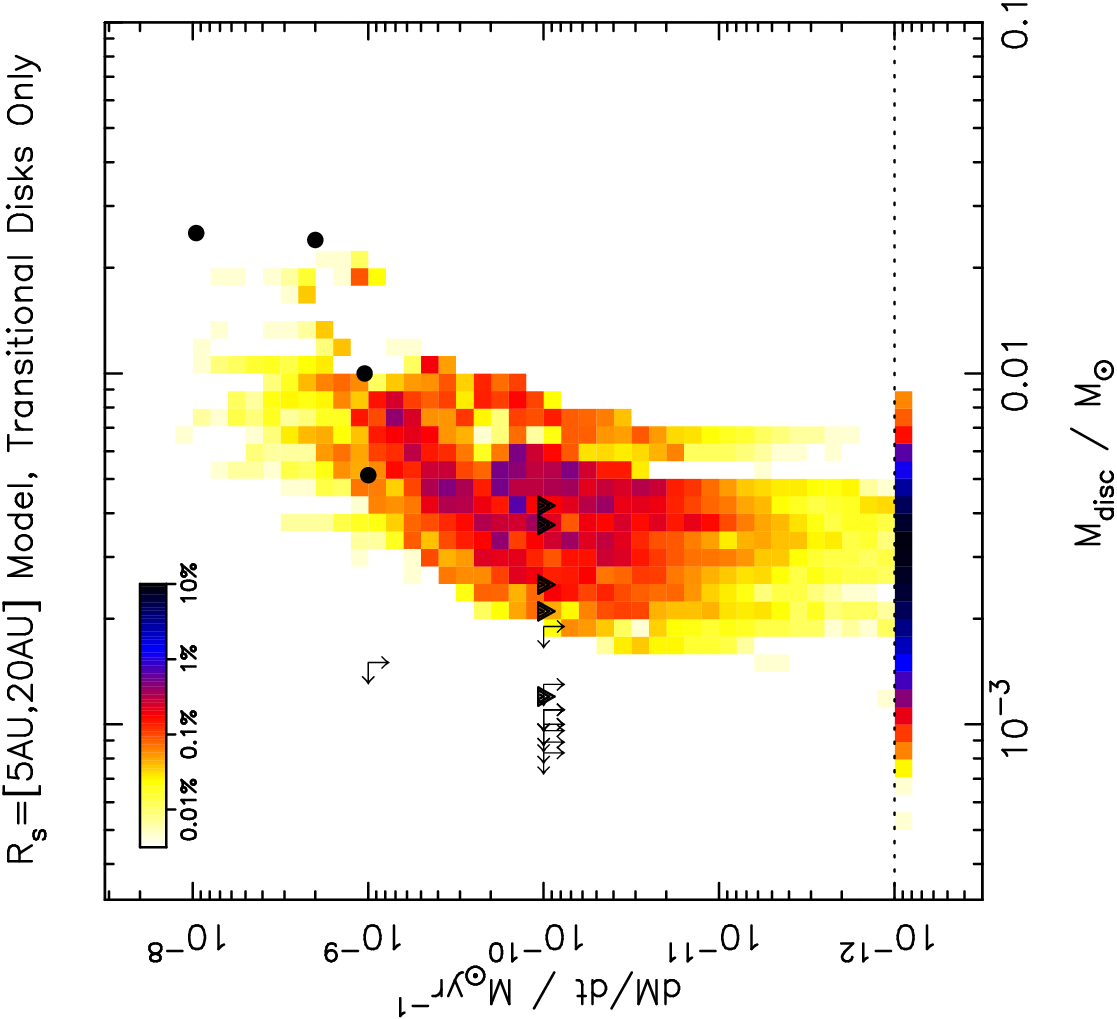}
\caption{Distributions of masses and accretion rates for transitional disks.  The upper panel shows the results from the {\sc reference} model set; the lower panel the {\sc scale} model set.  As in Fig.\ref{fig:mdot_time}, the pixels immediately below the dotted line represent upper limits.  All of the transition disks with accretion rates $>10^{-11}$\Msunyr\ contain planets.  Data points are taken from \citet[][circles]{najita07} and \citet[][single and double upper limits]{cieza08}, with the samples cut as described in the text.}\label{fig:mdot_mdisc}
\end{figure}

Both \citet{aa07} and \citet{najita07} suggested that the distribution of transitional disks in the $M_{\mathrm {disk}}$--${\dot{M}}$ plane can be a valuable diagnostic of the properties of transitional disks.  Fig.\ref{fig:mdot_mdisc} shows the distribution of accretion rates as a function of instantaneous disk mass for the transitional disks in the {\sc reference} and {\sc scale} model sets.  Also plotted are data from the studies of \citet{najita07} and \citet{cieza08}.  As our model has $M_* \equiv 1$\Msun, we only plot objects of spectral type M1 and earlier (corresponding to stars with $M_*\gtrsim0.5$\Msun).  We also omit known binaries from the figure, as well as disks whose classification is uncertain (those classified ``C/T'' by \citealt{najita07}).  All but one of the remaining stars in the \citet{cieza08} sample are WTTs with no measured accretion rates; we assign upper limits of $10^{-10}$\Msunyr\ to these objects\footnote{One object, USco J161420.2-190648, has a measured H$\alpha$ equivalent width of 52\AA~ \citep{preibisch02} but no detected UV-excess, so we assign it an upper limit of $10^{-9}$\Msunyr.}.  We note also that the binary statistics of the \citet{cieza08} sample are not well known.  We find reasonable agreement between the predictions of our models and the data, but find that the {\sc reference} model fails to reproduce the transition disks with the highest disk masses.  This discrepancy is within the systematic errors associated with the observations, but is also in part an artefact of the relatively small dispersion in our {\sc reference} disk model set.  The {\sc scale} model set, in which the disk scale radius $R_s$ was allowed to vary, shows a larger dispersion in the $M_{\mathrm {disk}}$--${\dot{M}}$ diagram, and provides a better fit to the observed transitional disk population.

All of the accreting ($>10^{-11}$\Msunyr) transitional disks in our models contain planets, while the overwhelming majority of the non-accreting transitional disks are undergoing photoevaporative clearing.  This is consistent with the predictions of previous studies \citep{aa07,najita07}, but we find that these two populations are not as well-separated in the $M_{\mathrm {disk}}$--${\dot{M}}$ plane as previous work has suggested.  In particular, $\sim35$\% of the planet-bearing transitional disks in our models have accretion rates in the range $10^{-12}$--$10^{-10}$\Msunyr (i.e., are accreting at a rate below the sensitivity limit of current observations).  This suggests that more sensitive observations, with limits of $\lesssim 10^{-11}$\Msunyr\, will be necessary to distinguish cleanly between the populations of planet-bearing and clearing transitional disks in this manner.


\subsection{Discussion}\label{sec:discussion}
These models are the first to attempt to predict the relative numbers of different types of transitional disks, but we stress that we are not yet capable of modeling all of the necessary physical processes simultaneously.  In particular, our models make no attempt to account for dust settling or growth, both of which are known to reduce the disk opacity and can give rise to ``transitional'' spectral energy distributions \citep{dd04,dd05}.  In essence we consider only the subset of transitional disks which have gaps or holes in their gas disks, and consequently our models are not capable of spanning the full spectrum of transition disk properties.  We also note that our ``gas-only'' definition of ``transitional'' is rather imprecise; more realistic models of dust-gas coupling \citep[e.g.,][]{aa07} are required to make more detailed predictions.  As a related point, we urge caution when comparing data to models, particularly with regard to selection criteria.  All but the most conservative observational definitions of ``transitional'' do not limit themselves to only objects with inner holes, and while the objects in these less strictly-selected samples are clearly interesting, they cannot readily be compared with models of the type presented here.

We further note that the ``planet-bearing'' disks identified in our models are limited to disks containing gas giant planets, which are massive enough to open disk gaps and migrate in the Type II regime.  Many exoplanets of lower mass are now known \citep[e.g.,][]{udry_ppv}, and comparison to the Solar System suggests that terrestrial planets can also form on time-scales comparable to observed disk lifetimes \citep[see, e.g.,][and references therein]{nagasawa_ppv}.  Planets less massive than $\sim 0.5$\Mjup\ are unlikely to open gaps in their parent gas disks, but the presence of even a low-mass planet can cause significant perturbations to the dust distribution \citep{pm06}.  It therefore seems possible that disks undergoing terrestrial planet formation may also be classed as transitional, even in the absence of the more dramatic perturbations to the disk structure modeled here.  However, the migration of low mass planets and planetary cores is theoretically complex and not fully understood \citep[e.g.,][]{pap_ppv,pp09}, and is not constrained by current exoplanet data.  Consequently any extension of our models beyond the Type II migration regime would introduce significant theoretical uncertainties, and would be of questionable benefit in this context.

In addition, we point out that our models apply only to stars of approximately solar mass.  Disks around low-mass stars and brown dwarfs are now commonly observed \citep[e.g.][]{scholz06}, but our knowledge of their evolution is limited.  Moreover, interpretation of the infrared spectral energy distributions of disks around low-mass (M-type) stars is fraught with difficulty \citep{ercolano09}, and it is not at all clear whether such disks evolve in the same manner as their more massive counterparts.  Very little is known about disk lifetimes and masses in this regime, and the key physical processes (angular momentum transport, photoevaporation, planet formation) are essentially unconstrained.  Consequently, we make no attempt to extrapolate our results to stars of lower mass.  However, it seems likely that future observations will discover a large number of transitional disks around low-mass stars \citep[e.g.,][]{sicilia08,currie09}, and this represents an interesting avenue for future work.


\section{Summary}\label{sec:summary}
We have constructed models of planet migration in evolving protoplanetary disks, and used a Monte Carlo approach to model the evolution of populations of such disks.  The disks evolve subject to viscosity and photoevaporation by the central star, and giant planets form and undergo Type II migration.  Our model successfully reproduces the frequency and radial distribution of observed extra-solar planets, and also reproduces the observed accretion rates, disk masses and lifetimes of protoplanetary disks.

The relatively small uniform sample of observed exoplanets limits the extent to which our models can inform our understanding of planet migration, but we are able to draw several interesting conclusions.  We find that the addition of the ``direct'' photoevaporative wind results in a degree of coupling between planet formation and disk clearing, which has not been seen in previous models.  Consequently some accretion flow across the planetary orbit must occur, as otherwise is is impossible to ``strand'' migrating giant planets at radii $\lesssim 1.5$AU.  We also find that planetary accretion during the migration phase cannot explain the observed range of exoplanet masses, and therefore conclude that the planet formation process must result in a broad range of giant planet masses.  Lastly, we find that it is only possible for giant planets to survive if they form at relatively late times (unless Type II migration is dramatically suppressed).  At this point in our models the disk surface densities at radii of a few AU are low ($\lesssim10$g cm$^{-2}$), and forming giant planets in such a low-density environment may be challenging for current theories of planet formation.

In addition, our models allow us to make a number of predictions about the properties and evolution of the so-called transitional disks (more precisely, the subset of transitional disks with holes or gaps in their gas disks).  Our models successfully reproduce the observed transition disk fractions of $\sim10$\%, and are also able to explain the accretion rates and disk masses of observed samples of transition disks.  However, we find that the properties of this population evolve significantly with time.  We predict the existence of two populations of transitional disks: weakly accreting disks, which contain embedded planets, and non-accreting disks which are undergoing inside-out clearing.  At early times ($\lesssim 2$Myr) essentially all transitional disks are planet-bearing, but the fraction of transition disks which possess planets drops with time; at late times ($\gtrsim 6$Myr), the vast majority of transitional disks are being cleared by photoevaporation.  Future observations will result in much larger samples of transitional disks than are currently known, and should allow us to disentangle the competing processes of disk evolution and planet formation.

\acknowledgments We are grateful to Adam Kraus for useful discussions about observations of young binaries, and to Barbara Ercolano for comments on the manuscript.  We thank Eric Mamajek \& Lucas Cieza for providing the data used in Figs.\ref{fig:disc_frac} \& \ref{fig:mdot_mdisc} respectively.  In addition we thank the referee, Scott Kenyon, for a thoughtful and insightful report.  We also acknowledge the generous hospitality of the Isaac Newton Institute for Mathematical Sciences, Cambridge, where this work was completed.  RDA acknowledges support from the Netherlands Organisation for Scientific Research (NWO), through VIDI grants 639.042.404 and 639.042.607.  PJA acknowledges support from the NSF (AST--0807471), from NASA's Origins of Solar Systems program (NNX09AB90G), and from NASA's Astrophysics Theory program (NNX07AH08G).


\end{document}